\documentclass[nofootinbib,showpacs,preprintnumbers,amsmath,amssymb,floatfix]
{revtex4-1}
\usepackage{graphicx}
\usepackage{epsf}
\usepackage{wrapfig}

\begin{document}

%\begin{center}

\title{Combining the small-x evolution and DGLAP for description of inclusive photon induced processes}

\vspace*{0.3 cm}

\author{B.I.~Ermolaev}
\affiliation{Ioffe Physico-Technical Institute, 194021
 St.Petersburg, Russia}
\author{S.I.~Troyan}
\affiliation{St.Petersburg Institute of Nuclear Physics, 188300
Gatchina, Russia}

\begin{abstract}
We study amplitudes of the inclusive photon induced high-energy processes:
elastic Compton scattering off hadrons and  photoproduction. Although description of these amplitudes includes both non-perturbative
and perturbative contributions, QCD factorization makes possible to study them separately. Throughout the present paper we focus on the perturbative
amplitudes and study such amplitudes in all available in the literature forms of QCD factorization:
Collinear, $K_T$ and Basic.
As a result, we obtain expressions for the perturbative Compton amplitudes, which can be used
at arbitrary $x$ and $Q^2$ in any form of QCD factorization. Putting $Q^2 = 0$ in those expressions allows us to
obtain expressions for the perturbative components of the photoproduction amplitudes.
The small-$x$ asymptotics of the Compton amplitudes in any form of QCD factorization are of the Regge type, with the Reggeon being a double-logarithmic
(non-BFKL) contribution to Pomeron.
\end{abstract}

\pacs{12.38.Cy}

\maketitle

\section{Introduction}

Photon induced inclusive processes such as Deep-Inelastic Scattering (DIS), Photoproduction and Diffractive Deep-Inelastic Scattering (DDIS) are the objects of
intensive experimental and theoretical investigation. Theoretical studying these processes
%involves QCD in both perturbative and non-perturbative domains and
is based on the concept of QCD factorization which makes possible to separate  perturbative and non-perturbative contributions.
There are several forms/types of QCD factorization
in the literature. Firstly, there is Collinear Factorization\cite{colfact}. Secondly, there is
the more general $K_T$ Factorization suggested in Refs.~\cite{ktfact,hefact}.
The third, the most general, type of QCD factorization is Basic Factorization
suggested in Ref.~\cite{bfact}. In any of those factorizations
the approximation of the single-parton
photon-hadron interactions is used and as a result, the subjects of consideration
%the DIS hadronic tensor $W_{\mu\nu}$ is
are represented through convolutions of perturbative and non-perturbative objects
connected by two-parton intermediate states. For instance, representation of
the amplitude $T$ of the elastic forward Compton scattering off a hadron
in any type of QCD factorization
%, related by Optical theorem to the structure function $F_1$,  is represented through convolutions
is depicted in Fig.~1 .

%%%%%%%%%%%%%%%%%%%%%%%%%%%%%%%%%%%%%%%%%%%%%%%%
\begin{figure}[h]
  \includegraphics[width=.7\textwidth]{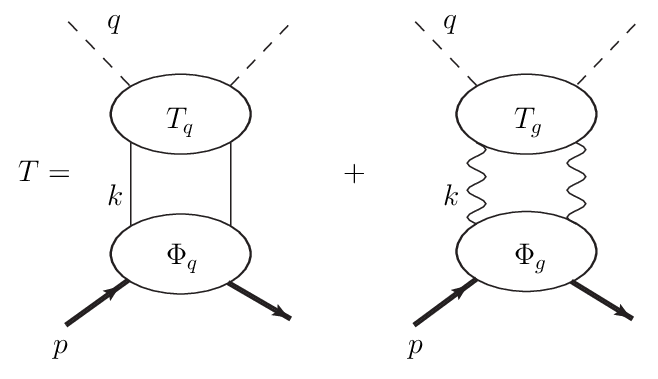}
  \caption{\label{combfig1} QCD factorization of amplitude T. The dashed lines denote virtual photons with momentum $q$, the horizontal straight lines
  correspond to the incoming hadron with momentum $p$, the vertical straight lines stand
  for intermediate quarks and the wavy lines denote intermediate gluons with momentum $k$. The upper blobs denote DIS off partons,
  they are described with Perturbative QCD while the lowest
  blobs are non-perturbative. }
\end{figure}
%%%%%%%%%%%%%%%%%%%%%%%%%%%%%%%%%%%%%%%%%%%%%%%%%%%

In the analytical form, QCD factorization of $T$ is generically written as follows:

\begin{equation}\label{fact}
T = T_q \otimes \Phi_q + T_g \otimes \Phi_g,
\end{equation}
where $T_q$ and $T_g$ are the perturbative amplitudes of the elastic Compton scattering off a quark and gluon respectively.
Non-perturbative blobs
$\Phi_q$ and  $\Phi_g$ denote  distributions of those partons in the hadrons. The notations $\otimes$ refer to integrations over momentum
$k$ of any parton in
the two-quark and two-gluon intermediate states. The specific form of the integrations as well as a parametrization of momentum $k$ of
the intermediate partons
 depend on the type of chosen
factorization\footnote{See for detail Ref.~\cite{bfact} and refs therein.}.  Amplitudes $T_{q,g}$
as well as amplitudes $\Phi_{q,g}$ are also given by expressions depending on the type of factorization. In the present paper we focus on the perturbative amplitudes $T_{q,g}$.
In any specific type of factorization, there  are kinematic regions where  amplitudes $T_{q,g}$ are given by essentially different expressions.\\
Such regions are
:\\
Region \textbf{A.} Large $x$  and large $Q^2$:  $(x \sim 1) \bigotimes (Q^2 \gg \mu^2)$.\\
Region \textbf{B.} Small $x$  and large $Q^2$: $(x \ll 1)\bigotimes (Q^2 \gg \mu^2)$.\\
Region \textbf{C.} Small $x$  and small $Q^2$: $(x \ll 1)\bigotimes(Q^2 < \mu^2)$.\\
Region \textbf{D.} Large $x$  and small $Q^2$: $(x \sim 1)\bigotimes(Q^2 < \mu^2)$.\\

We have used above the conventional notations: $Q^2 = - q^2$, with $q$ being the external photon momentum,
and $x = Q^2/w$, where $w = 2pq$, with $p$ being the momentum of the incoming/outgoing parton. The parameter
$\mu$ is associated with the factorization scale.
%, when Collinear Factorization is used. In addition, $\mu$
Besides, it plays the role of the infrared (IR)
cut-off in Collinear, $K_T$ and Basic Factorizations, when the double-logarithmic (DL) contributions to amplitudes $T_{q,g}$ are
accounted for. The physical meaning of $\mu$
for the parton distributions $\Phi_{q,g}$ in $K_T$ Factorization and its role in transition from $K_T$ Factorization to Collinear Factorization
were considered in Ref.~\cite{bfact}.
Now let us consider the status of knowledge of amplitudes $T_{q,g}$ in the Regions \textbf{A-D}.\\
The region \textbf{A} is the DGLAP applicability region, so expressions for $T_{q,g}$ in this region
are provided by the DGLAP evolution equations\cite{dglap}.
Such expressions can easily be found in the literature. They account for logarithms of $Q^2$ but neglect total resummations of logarithms of
$x$ because such logarithms are unessential in Region \textbf{ A}.\\
Description of $T_{q,g}$ in the small-$x$ region \textbf{B} in the double-logarithmic approximation (DLA) in the
framework of Collinear Factorization was obtained in Ref.~\cite{etf1}. In particular, it was shown in Ref.~\cite{etf1}
that the small-$x$ asymptotics of $T_{q,g}$ exhibits a new, DL contribution to Pomeron.
In contrast, double-logarithmic expressions for  $T_{q,g}$ in region \textbf{B } in $K_T$ Factorization
have not been obtained. We do it in the present paper, using the same method as in Ref.~\cite{etf1}: constructing and solving Infrared
Evolution Equations (IREE)\footnote{This method was suggested by L.N.~Lipatov. History of this method and its development are discussed in detail in
Ref.~\cite{egtg1sum}.}. Then we obtain expressions for  $T_{q,g}$
which combine DL and non-DL contributions available in DGLAP, so these expressions can
universally be used at large $Q^2$ and arbitrary $x$ (i.e. in the region $\textbf{A}\oplus \textbf{B}$) in Collinear, $K_T$
and Basic Factorizations. \\
Extending expressions for $T_{qg}$ to low $Q^2$ was suggested in Ref.~\cite{egtsmallq} for
Collinear Factorization. In the present paper we generalize this extension to the cases of $K_T$
and Basic Factorizations, obtaining thereby expressions which can universally be used at arbitrary $x$ and $Q^2$ in
any form of QCD factorization.\\
Putting $q^2 = 0$ in Eq.~(\ref{fact}), one arrives at the photoproduction amplitudes $A_{\gamma}$:

\begin{equation}\label{agammat}
A_{\gamma} \equiv T|_{q^2 = 0}.
\end{equation}

Combining it with Eq.~(\ref{agammat}) leads to the factorized form of $A_{\gamma q}$:
\begin{equation}\label{afact}
A_{\gamma} = A_{\gamma q} \otimes \Phi_q + A_{\gamma g} \otimes \Phi_g.
\end{equation}

The Optical theorem relates the amplitude $A_{\gamma}$
to the total cross section of photoproduction.
Using the expressions for $T_{qg}$ at low $Q^2$ allows us to obtain
expressions for the perturbative amplitudes $A_{\gamma q}, A_{\gamma g}$ first in Collinear and then in the
other forms of QCD factorization.

Our paper is organized as follows: in Sect.~II we remind results of Ref.~\cite{etf1} for
amplitudes $T_{q,g}$ in Region \textbf{B} in Collinear Factorization and
extend $T_{q,g}$ to Regions \textbf{A,C,D},
obtaining explicit expressions for $T_{q,g}$ which can be used at any $x$ and $Q^2$.
Then we briefly discuss the small-$x$ asymptotics of $T_{q,g}$, their applicability region
the power $Q^2$-corrections to $T_{q,g}$ in the low-$Q^2$ region \textbf{C.} In Sect.~III
we study amplitudes $T_{q,g}$ in $K_T$-Factorization. This type of factorization involves dealing with essentially off-shell
external partons, which brings a new technical problem which is absent in Collinear Factorization:
there is no universal description of $T_{q,g}$ in Region B at at arbitrary virtualities of the
external partons. We deal with this problem also by constructing and solving appropriate IREEs.  In Sect.~IV we consider the
photoproduction amplitudes in Collinear, $K_T$ and Basic Factorizations.
Finally, Sect.~V is for concluding remarks.

\section{Compton amplitudes in Collinear Factorization}

Eq.~(\ref{fact}) for Collinear Factorization takes the following form:

\begin{equation}\label{colfact}
T^{(col)} = \int_x^1 \frac{d \beta}{\beta}\left[ T^{(col)}_q(x/\beta,Q^2/\mu^2) \Phi^{(col)}_q (\beta,\mu^2)
+ T^{(col)}_g(x/\beta,Q^2/\mu^2) \Phi^{(col)}_g (\beta,\mu^2)\right],
\end{equation}
where the superscript $"col"$ refers to Collinear Factorization  and
$\beta$ is the longitudinal fraction of momentum $k$ in Fig.~1.
We start with considering amplitudes $T^{(col)}_{q,g}$
in region \textbf{B} and then extend these expressions to regions \textbf{A,C,D}.

%The photoproduction amplitudes $A_{\gamma q}$ and $A_{\gamma g}$ in Collinear Factorization do not depend on $Q^2$, so they are defined in two regions:\\
%Region \textbf{A}$^{\prime}$: Large $x$.\\
%Region \textbf{B}$^{\prime}$: Small $x$.\\

\subsection{Amplitudes $T^{(col)}_{q.g}$  in region B}

Perturbative amplitudes $T^{(col)}_{q,g}$
in region \textbf{B} were calculated in Ref.~\cite{etf1} in the double-logarithmic approximation (DLA) and the cases of fixed and running
$\alpha_s$ were investigated separately. Amplitudes $T^{(col)}_{q,g}$ were in Ref.~\cite{etf1} represented as follows:

%They were represented  in terms of the parton-parton amplitudes $h_{rr'}$:

\begin{eqnarray}\label{tqgcolb}
T^{(col)}_q (x/\beta, Q^2/\mu^2)&=& \int_{-\imath \infty}^{\imath \infty} \frac{d \omega}{2 \imath \pi} \left(x/\beta\right)^{-\omega}
\xi^{(+)} (\omega) F^{(col)}_q (\omega, y),
\\ \nonumber
T^{(col)}_g (x/\beta, Q^2/\mu^2)&=& \int_{-\imath \infty}^{\imath \infty} \frac{d \omega}{2 \imath \pi} \left(x/\beta\right)^{-\omega}
\xi^{(+)} (\omega) F^{(col)}_g (\omega,y),
\end{eqnarray}
with $\mu$ being the factorization scale.
 The logarithmic variable $y$ in Eq.~(\ref{tqgcolb}) is related to $Q^2$:
\begin{equation}\label{y}
y = \ln (Q^2/\mu^2)
\end{equation}
and $\xi^{(+)} (\omega)$ is the positive signature factor:
\begin{equation}\label{ksi}
\xi^{(+)} (\omega) = - \left(e^{-\imath \pi \omega} + 1\right)/2.
\end{equation}

The signature factor $\xi^{(+)} (\omega)$ guarantees that $T_{q,g}$ are invariant with respect to
permutation of the incoming and outgoing external photons. The integral representation (\ref{tqgcolb}) is
the asymptotic form of the Sommerfeld-Watson transform but it is frequently addressed
as the Mellin transform. Throughout the paper we will name $T_{q,g}$  the Mellin amplitudes.
Explicit expressions for the Mellin amplitudes $F^{(col)}_{q,g}$ in the region \textbf{B} are:
%at small-$x$ were obtained  in Ref.~\cite{etf1} where
%these amplitudes were represented in terms of the on-shell parton-parton amplitudes $A_{rr'}$ ($r,r' = q,g$) depicted in Fig.~2,
%and explicit expressions for them are presented in Appendix A.
%and $F^{(col)}_{q,g}$ defined as follows:

\begin{eqnarray}\label{fqgcolb}
F^{(col)}_q = C_q^{(+)} e^{\Omega_{(+)} y} + C_q^{(-)} e^{\Omega_{(-)} y},
\\ \nonumber
F^{(col)}_g = C_g^{(+)}  e^{\Omega_{(+)} y} +
C_g^{(-)}  e^{\Omega_{(-)} y}.
\end{eqnarray}

 Obtained in Ref.~\cite{etf1} expressions for $\Omega_{(\pm)}, C_q^{(\pm)}$ and $C_g^{(\pm)}$
can be found in Appendix B where they are expressed through the on-shell parton-parton amplitudes
$A_{rr'}$ ($r,r' = q,g$).  Graphs for amplitudes  $A_{rr'}$ are depicted in Fig.~2.

%%%%%%%%%%%%%%%%%%%%%%%%%%%%%%%%%%%%%%%%%%%%%%%%
\begin{figure}[h]
  \includegraphics[width=.7\textwidth]{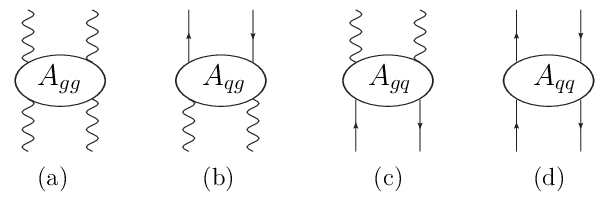}
  \caption{\label{combfig2}Graphs for the parton-parton amplitudes.}
\end{figure}
%%%%%%%%%%%%%%%%%%%%%%%%%%%%%%%%%%%%%%%%%%%%%%%%%%%

The Mellin transform for the parton-parton amplitudes is similar to the one in Eq.~(\ref{tqgcolb}):

\begin{equation}\label{mellinh}
A_{rr'} (s/\mu^2) = \int_{-\imath \infty}^{\imath \infty} \frac{d \omega}{2 \imath \pi} \left(s/\mu^2\right)^{\omega}
\xi^{(+)} (\omega) f_{rr'} (\omega).
\end{equation}

Technically, it is more convenient to use the Mellin amplitudes $h_{rr'}$ which are proportional to $f_{rr'}$:

\begin{equation}\label{fh}
h_{rr'}(\omega) = f_{rr'}(\omega)/(8 \pi^2).
\end{equation}

Explicit expression for amplitudes $h_{rr'}$ can be found in Appendix A.
Let us discuss Eq.~(\ref{fqgcolb}). It is easy to see that its structure  exhibits a distinct similarity to the
structure of
the DGLAP description of $F^{(col)}_{q,g}$, which is especially obvious when the approximation of fixed $\alpha_s$ is used:

\begin{eqnarray}\label{fqgdglap}
F^{DGLAP}_q &=& \hat{C}_q^{(+)} e^{\gamma_{(+)}^{DGLAP} y} + \hat{C}_q^{(-)} e^{\gamma_{(-)}^{DGLAP} y},
\\ \nonumber
F^{DGLAP}_g &=& \hat{C}_g^{(+)}  e^{\gamma_{(+)}^{DGLAP} y} +
\hat{C}_g^{(-)}  e^{\gamma_{(-)}^{DGLAP} y}.
\end{eqnarray}

 Indeed, the factors $\Omega_{(\pm)}$  in the exponents of  Eq.~(\ref{fqgcolb}) are new
anomalous dimensions instead of  $\gamma^{DGLAP}_{(\pm)}$ in Eq.~(\ref{fqgdglap}) while
the factors $C_g^{(\pm)}, C_q^{(\pm)}$ are new coefficient functions instead of the DGLAP coefficient functions $\hat{C}^{(\pm)}_{q,g}$.
However in contrast to DGLAP, the coefficient functions and the anomalous dimensions in (\ref{fqgcolb}) are calculated with the same means
and they include the total resummations of the DL contributions.
The both coefficient functions and anomalous dimensions in (\ref{fqgcolb}) are made of amplitudes $h_{rr'}$. When
the partonic amplitudes $h_{rr'}$ are replaced by their Born values, the integrands in Eq.~(\ref{tqgcolb}) coincide with the
integrands of the LO DGLAP expressions in which the most singular in $\omega$ terms are retained. Further expansions of
$h_{rr'}$ and substituting them in Eq.~(\ref{fqgcolb}) lead to the NLO (as well as to NNLO DGLAP, etc), where the most singular terms only
are accounted for in each order in $\alpha_s$.

\subsection{Extending the small-x Eqs.~(\ref{tqgcolb},\ref{fqgcolb}) to region A}

Expressions in Eqs.~(\ref{tqgcolb},\ref{fqgcolb}) are defined in the region \textbf{B}.
%Extending down to the region \textbf{A} is not as rigorous as
%obtaining (\ref{tqgcolb},\ref{fqgcolb}) where the regular method of constructing and solving IREE was applied.
Now we are going to obtain an interpolation formulae for  amplitudes $T^{(col)}_{q,g}$, which would
reproduce the DGLAP expressions at large $x$ and at the same time would coincide with Eqs.~(\ref{tqgcolb},\ref{fqgcolb})
at small $x$.
Such extension can be obtained  with the four steps: \\
\textbf{Step 1:} Subtract the most singular in $\omega$ terms (i.e. the terms $\sim \alpha_s/\omega$ in
the case of LO DGLAP; $\alpha^2_s/\omega^3$ in the case of NLO DGLAP, etc.) from the DGLAP anomalous dimensions $\gamma_{(\pm)}^{(DGLAP)} (\omega, \alpha_s)$. We
denote $\widetilde{\gamma}^{DGLAP}_{(\pm)}$ the result of such amputation. \\
\textbf{Step 2:} Add $\widetilde{\gamma}^{DGLAP}_{(\pm)}$ to $\Omega_{(\pm)}$.
%The result of such addition is denoted $\widetilde{h}_{rr'}$ in Eq,~(\ref{htilde}).
The new anomalous dimensions

\begin{equation}\label{omegatilde}
\widetilde{\Omega}_{(\pm)} = \Omega_{(\pm)} + \widetilde{\gamma}^{DGLAP}_{(\pm)}
\end{equation}

contain the total resummations of the double-logarithmic contributions. They are essential at small $x$
and unimportant at large $x$.
At the same time, (\ref{omegatilde}) contains less singular in $1/\omega^n$ DGLAP contributions which are dominant at large $x$.
\\
\textbf{Step 3:} Subtract the LO DGLAP term $=1$ and the most singular in $\omega$ terms (i.e. the term 1 $\sim \alpha_s/\omega^2$ in the case
of NLO DGLAP, $\alpha^2_s/\omega^4$ in the case of NNLO DGLAP, etc.) from the DGLAP expressions for the coefficient
functions. We denote $\Delta \hat{C}^{(\pm)}_{q,g}$ the result of such amputation.
\\
\textbf{Step 4:} Add the results obtained in Step 3 to the DL expressions $C^{(\pm)}_{q,g}$,
arriving thereby at new coefficient functions $\widetilde{C}^{(\pm)}_{q,g}$:

\begin{equation}\label{cqgtilde}
\widetilde{C}^{(\pm)}_{q,g} = C^{(\pm)}_{q,g} + \Delta \hat{C}^{(\pm)}_{q,g}.
\end{equation}

%As a result, the coefficient functions $\widetilde{C}^{(\pm)}_{q,g}$
The coefficient functions  $\widetilde{C}^{(\pm)}_{q,g}$  include both DL contributions and the less singular DGLAP terms.
The subtractions in Steps 1,3 are
necessary to avoid the double counting.

Replacing $\Omega_{(\pm)}$ and $C^{(\pm)}_{q,g}$  in Eq.~(\ref{tqgcolb}) by
 $\widetilde{\Omega}_{(\pm)}$ and $\widetilde{C}^{(\pm)}_{q,g}$, we obtain
 the interpolation expressions $\widetilde{T}^{(col)}_{q,g}$ for the Compton amplitudes :

\begin{eqnarray}\label{tqgcolbtilde}
\widetilde{T}^{(col)}_q (x/\beta, Q^2/\mu^2)&=& \int_{-\imath \infty}^{\imath \infty} \frac{d \omega}{2 \imath \pi} \left(x/\beta\right)^{-\omega}
\xi^{(+)} (\omega) \left[\widetilde{C}_q^{(+)} e^{\widetilde{\Omega}_{(+)} y} + \widetilde{C}_q^{(-)} e^{\widetilde{\Omega}_{(-)} y}\right],
\\ \nonumber
T^{(col)}_g (x/\beta, Q^2/\mu^2)&=& \int_{-\imath \infty}^{\imath \infty} \frac{d \omega}{2 \imath \pi} \left(x/\beta\right)^{-\omega}
\xi^{(+)} (\omega)\left[ \widetilde{C}_g^{(+)}  e^{\widetilde{\Omega}_{(+)} y} +
\widetilde{C}_g^{(-)}  e^{\widetilde{\Omega}_{(-)} y} \right].
\end{eqnarray}
%where we have denoted that  $\Omega_{(\pm)}$  and $C_{q,g}^{(\pm)}$  are replaced by $\widetilde{\Omega}_{(\pm)}$
%and $\widetilde{C}_{q,g}^{(\pm)}$  respectively when $h_{rr'}$ is replaced by $\widetilde{h}_{rr'}$.
Eq.~(\ref{tqgcolbtilde}) combines the small-$x$ evolution in DLA with the DGLAP results for the coefficient functions and anomalous dimensions,
which are important at large $x$. These expressions can universally be used as the interpolation formulae for $T^{(col)}_{q,g}$ in the region $\textbf{A}  \oplus  \textbf{B}$.

\subsection{Extending Eq.~(\ref{tqgcolbtilde}) to the small-$Q^2$ region $\textbf{C} \oplus\textbf{ D}$}

Extension of the expressions (\ref{tqgcolbtilde}) to describe the amplitudes
$\widetilde{T}^{(col)}_{q,g}$ to the region $\textbf{C} \oplus \textbf{D}$ is also not rigorous.
The standard approach suggested in Ref.~\cite{nacht} and used since that in many papers (see e.g. Ref.~\cite{bad}) to describe DIS
at low  $Q^2$ is to make a shift
of $Q^2$:

\begin{equation}\label{shift}
Q^2 \to Q^2 + m^2,
\end{equation}
where the mass scale $m$ was fixed on basis of certain physical considerations, depending on specific situation. In contrast to preceding papers,
we proved in Ref.~\cite{egtsmallq}
(see also the overview\cite{egtg1sum}) that the scale $m$ with DL accuracy
can be unambiguously specified: the IR cut-off $\mu$ plays the role of the scale $m$.
Our proof was based on the well-known fact that DL contributions arise from integrations
 $\sim d k^2_{i \perp}/k^2_{i \perp}$
 over the transverse momenta $k_{i \perp}$ of virtual partons, which
 requires introducing an IR cut-off $\mu$.
 It can be introduced through the shift $k^2_{i \perp} \to k^2_{i \perp} + \mu^2$, which
 eventually leads to the
 shift (\ref{shift}), with

 \begin{equation}\label{mum}
 m = \mu.
 \end{equation}

The proof of (\ref{mum}) in Refs.~\cite{egtsmallq,egtg1sum} was
done in the context of the spin-dependent structure function $g_1$ but it holds for amplitudes
$\widetilde{T}^{(col)}_{q,g}$ either.  Applying the Principle of Minimal Sensitivity\cite{pms},  we estimated
in Ref.~\cite{etf1} the value
of $\mu$ of Eq.~(\ref{mum}):

\begin{equation}\label{mulambda}
\mu = 2.3 \Lambda_{QCD} \approx 1 GeV.
\end{equation}

 So, the universal expression for the Compton amplitudes $\widetilde{T}^{(col)}_{q,g}$ valid in the region
 $\textbf{A} \oplus \textbf{B}\oplus  \textbf{C} \oplus \textbf{D}$
(i.e. at arbitrary values of $x$ and $Q^2$) in the framework of Collinear Factorization is

\begin{eqnarray}\label{tqgcol}
\widetilde{T}^{(col)}_q (x/\beta, Q^2/\mu^2)&=& \int_{-\imath \infty}^{\imath \infty} \frac{d \omega}{2 \imath \pi}
\left(\widetilde{x}/\beta\right)^{-\omega}
\xi^{(+)} (\omega) \left[\widetilde{C}_q^{(+)} e^{\widetilde{\Omega}_{(+)} \widetilde{y}} + \widetilde{C}_q^{(-)} e^{\widetilde{\Omega}_{(-)} \widetilde{y}}\right],
\\ \nonumber
\widetilde{T}^{(col)}_g (x/\beta, Q^2/\mu^2)&=& \int_{-\imath \infty}^{\imath \infty} \frac{d \omega}{2 \imath \pi} \left(\widetilde{x}/\beta\right)^{-\omega}
\xi^{(+)} (\omega)\left[ \widetilde{C}_g^{(+)}  e^{\widetilde{\Omega}_{(+)} \widetilde{y}} +
\widetilde{C}_g^{(-)}  e^{\widetilde{\Omega}_{(-)} \widetilde{y}} \right],
\end{eqnarray}
where we have denoted

\begin{equation}\label{xytilde}
\widetilde{x} = x + \mu^2/w \equiv x + x_0, ~~~~ \widetilde{y} = \ln \left((Q^2 + \mu^2)/\mu^2\right).
\end{equation}

\subsection{Small-$x$  asymptotics of the Compton amplitudes and comparison with the DGLAP asymptotics}

Asymptotics of $T^{(col)}_{q,g}$ at $x \to 0$ and $Q^2 > \mu^2$ were considered in detail in Ref.~\cite{etf1}. In the present paper we
briefly remind results of Ref.~\cite{etf1} and compare these asymptotics
with the asymptotics of the same amplitudes obtained in the DGLAP approach.
%There is no difference between the asymptotics of  $T^{(col)}_q$ and $T^{(col)}_g$.
The standard mathematical tool to
calculate the small-$x$ asymptotics of $T^{(col)}_{q,g}$ is Saddle Point method. Applying this method to
$T^{(col)}_{q,g}$, we obtain that the small-$x$ asymptotics of
amplitudes $T^{(col)}_{q,g}$ are of the Regge type. They both have the same intercepts $\omega_0$:

\begin{eqnarray}\label{asympt}
T^{(col)}_q &\sim&  \frac{\Pi_q(\omega_0)}{(\ln (1/x))^{3/2}} x^{- \omega_0} (Q^2/\mu^2)^{\omega_0/2},~~
T^{(col)}_g \sim  \frac{\Pi_g(\omega_0)}{(\ln (1/x))^{3/2}} x^{- \omega_0} (Q^2/\mu^2)^{\omega_0/2}.
\end{eqnarray}
The factors $\Pi_{q,g}(\omega_0)$ in Eq.~(\ref{asympt}) are called the impact factors\footnote{Explicit expressions
 for the impact factors can be found un Ref.~\cite{etf1}.} They are the only difference between the
 asymptotics of $T^{(col)}_q $ and $T^{(col)}_g$.
 The value of the intercept $\omega_0$ in Eq.~(\ref{asympt}) proved to be dependent on accuracy of the calculations.
%was calculated in Ref.~\cite{etf1} under several different approximations.
The maximal intercept
$\omega^{DL}_H$ corresponded to the roughest approximation where the quark contributions were
neglected and $\alpha_s$ was fixed while the minimal intercept  $\omega^{DL}_S$
%was close to the NLO BFKL Pomeron intercept\cite{nlobfkl}. It
corresponded to the most accurate calculation where both gluon and quark contributions were accounted for
and $\alpha_s$ was running:

%This intercept was close to the LO BFKL Pomeron intercept\cite{lobfkl} These intercepts are:

\begin{equation}\label{inths}
\omega^{DL}_H = 1.35,~~\omega^{DL}_S = 1.07.
\end{equation}

As the both intercepts $> 1$, the Reggeons of Eq.~(\ref{asympt}) with the intercepts
(\ref{inths}) are the DL contributions to Pomeron, or DL Pomerons.
In accordance with the conventional Pomeron terminology we call them  hard (with the subscript H)
 and soft (with the subscript S) DL Pomerons respectively.
%Eq.~(\ref{inths}) reads that the Reggeon in (\ref{asympt}) turns out to be a supercritical Pomeron.
It is interesting fact that $\omega^{DL}_H$ is close to the LO BFLK Pomeron intercept and
$\omega^{DL}_S$ almost coincides with the NLO BFLK Pomeron intercept despite that BFKL have nothing in common with
resummation of DL contributions.
The asymptotics in Eq.~(\ref{asympt}) are given by much simpler expressions than the parent amplitudes in Eq.~(\ref{tqgcol}).
However, the asymptotics should not be used outside their applicability region. It was shown in Ref.~\cite{etf1} that
the asymptotics should be used at $x < x_{max}$ and estimates of $x_{max}$ at various values of $Q^2$ were obtained.
%This region depends on .  In particular,
%Ref.~\cite{etf1} states that
In particular, at $Q^2 = 10$~GeV$^2$
%the asymptotics (\ref{asympt}) reliably represent the amplitudes (\ref{tqgcol}) at $x \leq x_{max}$ only, with

\begin{equation}\label{xmax}
x_{max} = 10^{-6}.
\end{equation}
It was shown in Ref.~\cite{etf1} that the greater is $Q^2$, the less is $x_{max}$. When $x \geq x_{max}$ the parent amplitudes
$T_{q,g}$ should be used instead of the asymptotics.

\subsection{Remark on Asymptotic Scaling}

Eq.~(\ref{asympt}) can be written in such a way:

\begin{equation}\label{asymptzeta}
T^{(col)}_q \sim \frac{\Pi_q(\omega_0)}{\ln^{3/2} (1/x)}
\left(\zeta/\mu^2\right)^{\omega_0/2},~~T^{(col)}_g \sim \frac{\Pi_g(\omega_0)}{\ln^{3/2} (1/x)}
\left(\zeta/\mu^2\right)^{\omega_0/2},
\end{equation}
with

\begin{equation}\label{zeta}
\zeta = Q^2/x^2.
\end{equation}

Although both $T^{(col)}_q$ and $T^{(col)}_g$ by definition depend on two independent variables $x$ and $Q^2$,
Eq.~(\ref{asymptzeta}) manifests that
  $T^{(col)}_{q,g} \ln^{3/2} (1/x)$ asymptotically depend
on the single variable $\zeta$ only. We name  this remarkable property the Asymptotic Scaling.
%It is the consequence of accounting for DL contributions to all orders in $\alpha_s$.
This property is
absent in the DGLAP approach. Indeed, DGLAP predicts that the $x$- and $Q^2$- dependence of   $T^{(col)}_{q,g}$
at $x \to 0$ are
unrelated:

\begin{equation}\label{asdglap}
T^{(DGLAP)}_{q,g} \sim x^{-a} (Q^2/\mu^2)^{\gamma_{DGLAP}/b},
\end{equation}
where $\gamma_{DGLAP}$ is the anomalous dimension and $b$ is the first coefficient of the $\beta$-function.
The intercept $a$ has the phenomenological origin: it is generated by the terms $x^{-a}$ in the fits for the
initial parton densities.

\subsection{Perturbative power $Q^2$-contributions}

We begin studying the power $Q^2$-corrections with considering
$T^{(col)}_q$ at the large-$Q^2$ region, where $Q^2 \gg \mu^2$ and where the logarithmic variable $\widetilde{y}$
defined in (\ref{xytilde}) can be expanded in the series as follows:

\begin{equation}\label{ybigqser}
\widetilde{y} = y + \sum_{n=1}^{\infty} c_n \left(\frac{\mu^2}{Q^2}\right)^n.
\end{equation}

Being substituted in Eq.~(\ref{tqgcol}), it allows us to write $T^{(col)}_q$ as

\begin{equation}\label{tbigqser}
\widetilde{T}^{(col)}_q (x/\beta, \widetilde{Q}^2) = \widetilde{T}^{(col)}_q (x/\beta,Q^2) +
\left(\frac{\mu^2}{Q^2}\right) C_1  +
\left(\frac{\mu^2}{Q^2}\right)^2  C_2  +...
\end{equation}

The first term in the r.h.s. of Eq.~(\ref{tbigqser}) describes the Compton amplitude in the region $\textbf{A}\otimes\textbf{B}$.
At large $x$ it is given by the DGLAP formulae whereas
 the second and third terms  can wrongly be interpreted
as contributions of the higher twists, probably of the non-perturbative origin
 though actually they are altogether perturbative:

\begin{equation}\label{c12}
C_1 = \frac{\partial \widetilde{T}^{(col)}_q (x/\beta,y)}{\partial y},~~
C_2 = \frac{1}{2} \frac{\partial^2 \widetilde{T}^{(col)}_q (x/\beta,y)}{\partial y^2}.
\end{equation}

This example demonstrates that ignoring the shift (\ref{shift}) and interpretating all terms inversely
proportional to $Q^2$ as an impact of higher twist could be misleading. On the other,
at $Q^2 < \mu^2$ the expansion (\ref{ybigqser}) is replaced by another one:

\begin{equation}\label{ysmallser}
\widetilde{y} =  \sum_{n=1}^{\infty} c'_n \left(\frac{Q^2}{\mu^2}\right)^n.
\end{equation}

It leads to the small-$Q^2$ expression for $\widetilde{T}^{(col)}_q$:

\begin{equation}\label{tsmallqser}
\widetilde{T}^{(col)}_q (x/\beta, \widetilde{Q}^2)= \sum_{n-1}^{\infty}  \left(\frac{Q^2}{\mu^2}\right)^n C'_n
\end{equation}
which decreases down to zero when $Q^2 \to 0$.  It explains very well why the terms $\sim 1/(Q^2)^n$
are never seen at small $Q^2$ where, naively, their impact could be
great.  We remind that our estimate of $\mu$ is $\mu \approx 1$~GeV (see Eq.~(\ref{mulambda})).
Disappearing of the terms $\sim 1/(Q^2)^n$ at $Q^2 \lesssim 1$~GeV  obtained from
analysis of experimental data could confirm correctness of our reasoning.

\section{ Calculating perturbative Compton amplitudes in $K_T$ Factorization}

Originally, $K_T$ Factorization was introduced in Refs.~\cite{ktfact,hefact} to  make possible applying BFKL for
description of hadronic reactions. It implied the kinematic region of
very high, or asymptotic, energies, i.e. at $x \ll 1$. However, in the cases when BFKL is not involved one can
%use either Collinear Factorization or
use $K_T$ Factorization at $x \sim 1$ as well.  In the present Sect. we generalize DL
expressions for $T^{(col)}_{q,g}$ to  $K_T$ Factorization.
In the framework of $K_T$ Factorization Eq.~(\ref{fact}) takes the following form:

\begin{equation}\label{ktfact}
T^{(KT)} = \int_x^1 \frac{d \beta}{\beta}\int_{\mu^2}^w \frac{d k^2_{\perp}}{k^2_{\perp}}\left[ T^{(KT)}_q(x/\beta,Q^2,k^2_{\perp},\mu^2) \Phi^{(KT)}_q (\beta,k^2_{\perp},\mu^2)
+ T^{(KT)}_g(x/\beta,Q^2,k^2_{\perp},\mu^2) \Phi^{(KT)}_g (\beta,k^2_{\perp},\mu^2)\right],
\end{equation}
where the superscript $"KT"$ refers to $K_T$ -Factorization, $\mu$ is the factorization scale,
$\beta$ is the longitudinal fraction of the external partons (see Fig.~1) and $k^2_{\perp}$ stands for their transverse momentum.
$\Phi^{(KT)}_{q,g}$ denote initial parton distributions and amplitudes $T^{(KT)}_{q,g}$ are perturbative.
The subscripts $"q,g"$ refer to quarks and gluons in the same way as in Sect.~II.
 We are going to calculate $T^{(KT)}_{q,g}$  in DLA. DL contributions to them
arrive from the kinematic region

\begin{equation}\label{wqk}
w \gg Q^2, k^2_{\perp} \gg \mu^2.
\end{equation}

However, amplitudes $T^{(KT)}_{q,g}$ in DLA cannot be described  by a single universal expression
at different virtualities $k^2_{\perp}$ in (\ref{wqk}).
 There are two regions, where they are given by different expressions: \\
Region \textbf{E}: Moderate-virtual $k^2_{\perp}$,  where

 \begin{equation}\label{mvkin}
 k^2_{\perp} \ll \mu^2/x.
 \end{equation}
Region \textbf{F}: Deeply-virtual $k^2_{\perp}$, where

 \begin{equation}\label{dvkin}
   k^2_{\perp} \gg \mu^2/x.
 \end{equation}

In order to avoid confusions, we will denote $T^{(\textbf{E})}_{q,g}$ and $T^{(\textbf{F})}_{q,g}$ the
amplitudes $T^{(KT)}_{q,g}$ in the regions \textbf{E }and \textbf{F} respectively.

We will use the Mellin transform for $T^{(\textbf{E,F)}}_{q,g}$ in the following form:

\begin{eqnarray}\label{mellinphi}
T^{(\textbf{E})}_{q,g} = \int_{-\imath \infty}^{\imath \infty} \frac{d \omega}{2 \imath \pi} \left(w/\mu^2\right)^{\omega}
\xi^{(+)} (\omega) \varphi^{(\textbf{E})}_{q,g}(\omega, y, z),
\\ \nonumber
T^{(\textbf{F})}_{q,g} = \int_{-\imath \infty}^{\imath \infty} \frac{d \omega}{2 \imath \pi} \left(w/\mu^2\right)^{\omega}
\xi^{(+)} (\omega) \varphi^{(\textbf{F})}_{q,g}(\omega, y, z),
\end{eqnarray}
where $y$ is given by Eq.~(\ref{y}) and the new variable

\begin{equation}\label{z}
z = \ln (k^2_{\perp}/\mu^2)
\end{equation}
describes dependence of $T^{(KT)}_{q,g}$ on $k_{\perp}$ in the kinematics (\ref{wqk}).
This dependence differs
%$k_{\perp}^2 = \mu^2$, then
$T^{(KT)}_{q,g}$  from  amplitudes $T^{(col)}_{q,g}$ considered in Sect.~II in Collinear Factorization.
In addition to $y$ and $z$, it is convenient to introduce one more logarithmic variable:

\begin{equation}\label{rho}
\rho = \ln (w/\mu^2).
\end{equation}

In terms of the logarithmic variables regions \textbf{E} and \textbf{F}
defined in Eqs.~(\ref{mvkin},\ref{dvkin}) look as follows: \\
Region \textbf{E}: $\rho > y + z$,\\
Region \textbf{F}: $\rho < y + z$. \\

\subsection{Calculating the off-shell Compton amplitudes in region E}

IREEs for $\varphi^{(\textbf{E})}_{q,g}$ have the following form:

\begin{eqnarray}\label{eqphikt}
\frac{\partial \varphi^{(\textbf{E})}_q }{\partial y} + \frac{\partial \varphi^{(\textbf{E})}_q }{\partial z}  +
\omega \varphi^{(\textbf{E})}_q   = F^{(col)}_q (\omega,y) H_{qq} (\omega,z)
+ F^{(col)}_g (\omega,y) H_{gq} (\omega,z),
\\ \nonumber
\frac{\partial \varphi^{(\textbf{E})}_g}{\partial y} +\frac{\partial \varphi^{(\textbf{E})}_g}{\partial z} +
\omega \varphi^{(\textbf{E})}_g
= F^{(col)}_q (\omega, y) H_{qg} (\omega,z)
+ F^{(col)}_g (\omega,y) H_{gg} (\omega,z).
\end{eqnarray}

The l.h.s. of IREEs in (\ref{eqphikt}) are obtained with applying the differential operator

\begin{equation}\label{dmu}
-\mu^2  d/d \mu^2 = \partial/\partial \rho + \partial/\partial y + \partial/\partial z
\end{equation}
 to Eq.~(\ref{mellinphi}).
As a result, the l.h.s.of Eq.~(\ref{eqphikt}) contain explicitly the derivatives over $y$ and $z$ while the factor $\omega$ corresponds to
differentiation of the Mellin factor $(w/\mu^2)^{\omega}$.
The convolutions in r.h.s. of  (\ref{eqphikt}) are similar in structure to the convolutions in the DGLAP evolution equations.
They involve amplitudes $F^{(col)}_q$ which were considered in Sect.~II and
 amplitudes $H_{rr'}$ which are still unknown.
% They correspond to the parton-parton amplitudes $A'_{rr'}$, where  the lower partons (see Fig.~2) are off-shell.
 We will calculate amplitudes $H_{rr'}$ in the next Sect.
In order to specify general solutions to Eq.~(\ref{eqphikt}), we will use the matching

\begin{equation}\label{matche}
T^{(\textbf{E})}_q (w,y,z)|_{z = 0} = T^{(col)}_q (\omega, y),~~~
T^{(\textbf{E})}_g(w,y,z)|_{z = 0} = T^{(col)}_g (\omega, y),
\end{equation}
with $T^{(col)}_{q,g}$ defined in Eqs.~(\ref{tqgcolb}).
The matching (\ref{matche}) implies the following hierarchy between $y$ and $z$:

\begin{equation}\label{yz}
 y > z.
\end{equation}

The requirement (\ref{yz}) is temporary. The final expressions for $T^{(\textbf{E})}_{q,g}$ will be written in the
form free of this hierarchy.
Solving Eq.~(\ref{eqphikt}) goes easier when $y$ and $z$ are replaced by new variables  $\xi, \eta$:

\begin{equation}\label{ksiy}
\xi = \frac{1}{2} (y + z),~~~\eta =  \frac{1}{2} (y - z).
\end{equation}

Obviously, the inverse transform is

\begin{equation}\label{yksi}
y = \xi + \eta,~~~z = \xi - \eta.
\end{equation}

In terms of $\xi,\eta$ Eq.~(\ref{eqphikt}) looks simpler:

\begin{eqnarray}\label{eqphiksi}
\frac{\partial \varphi^{(\textbf{E})}_q }{\partial \xi} +
\omega \varphi^{(\textbf{E})}_q   = F^{(col)}_q (\omega,y) H_{qq} (\omega,z)
+ F^{(col)}_g (\omega,y) H_{gq} (\omega,z),
\\ \nonumber
\frac{\partial \varphi^{(\textbf{E})}_g}{\partial \xi}  +
\omega \varphi^{(\textbf{E})}_g
= F^{(col)}_q (\omega, y) H_{qg} (\omega,z)
+ F^{(col)}_g (\omega,y) H_{gg} (\omega,z).
\end{eqnarray}

Solution to Eq.~(\ref{eqphiksi}) is:

\begin{eqnarray}\label{phie}
\varphi^{(\textbf{E})}_q = e^{-\omega \xi} \left[T^{(col)}_q (\omega, \eta)  +
\int_{\eta}^{\xi} d v e^{\omega v}\left(F^{(col)}_q (\omega,y') H_{qq} (\omega,z')
+ F^{(col)}_g (\omega,y') H_{gq} (\omega,z')\right)\right],
\\ \nonumber
\varphi^{(\textbf{E})}_g = e^{-\omega \xi} \left[T^{(col)}_g (\omega, \eta)  +
\int_{\eta}^{\xi} d v e^{\omega v}\left(F^{(col)}_q (\omega,y') H_{qg} (\omega,z')
+ F^{(col)}_g (\omega,y') H_{gg} (\omega,z')\right)\right],
\end{eqnarray}
with the variables $y',z'$ defined as follows:

\begin{equation}\label{yprime}
y' = v + \eta,~~z' = v - \eta.
\end{equation}

 In order to lift the relation between $y$ and $z$ of
Eq.~(\ref{yz}) we replace $\eta$ with $|\eta|$ everywhere save Eq.~(\ref{yprime}).
Doing so and
substituting Eq.~(\ref{phie}) in (\ref{mellinphi}), we arrive at expressions for amplitudes
$T^{(\textbf{E})}_{q,g}$ in region \textbf{E}:

\begin{eqnarray}\label{tkte}
T^{(\textbf{E})}_q = \int_{-\imath \infty}^{\imath \infty} \frac{d \omega}{2 \imath \pi}
\xi^{(+)} (\omega) \left(\frac{s}{\sqrt{Q^2 k^2_{\perp}}}\right)^{\omega}
\left[T^{(col)}_q (\omega, |\eta|)  +
\right.
\\ \nonumber
\left.
\int_{|\eta|}^{\xi} d v e^{\omega v}\left(F^{(col)}_q (\omega,y') H_{qq} (\omega,z')
+ F^{(col)}_g (\omega,y') H_{gq} (\omega,z')\right)\right],
\\ \nonumber
T^{(\textbf{E})}_g = \int_{-\imath \infty}^{\imath \infty} \frac{d \omega}{2 \imath \pi}
\xi^{(+)} (\omega) \left(\frac{s}{\sqrt{Q^2 k^2_{\perp}}}\right)^{\omega}
\left[T^{(col)}_g (\omega, |\eta|)  +
\right.
\\ \nonumber
\left.
\int_{|\eta|}^{\xi} d v e^{\omega v}\left(F^{(col)}_g (\omega,y') H_{qg} (\omega,z')
+ F^{(col)}_g (\omega,y') H_{gg} (\omega,z')\right)\right].
\end{eqnarray}

\subsection{Calculating the off-shell Compton amplitudes in region F}

%Integrations over momenta of all virtual partons, including the softest partons, in region \textbf{F}
%do not involve the cut-off $\mu$, so
%the r.h.s. of the both IREEs for $T^{(\textbf{F})}_{q,g}$ is zero which makes such
IREEs for $T^{(\textbf{F})}_{q,g}$  are very simple:

\begin{eqnarray}\label{eqtktf}
\frac{\partial T^{(\textbf{F})}_q}{\partial \rho} + \frac{\partial T^{(\textbf{F})}_q}{\partial y}
+ \frac{\partial T^{(\textbf{F})}_q}{\partial z} = 0,
\\ \nonumber
\frac{\partial T^{(\textbf{F})}_g}{\partial \rho} + \frac{\partial T^{(\textbf{F})}_g}{\partial y}
+ \frac{\partial T^{(\textbf{F})}_g}{\partial z} = 0.
\end{eqnarray}

We do not use the Mellin transform for  $T^{(\textbf{F})}_{q,g}$ and because of it the l.h.s. of (\ref{eqtktf})
contain derivatives over $\rho$, $y$ and $z$. Integration over momenta of all
virtual partons in the deeply-virtual region \textbf{F} do not bring any dependence\footnote{A detailed derivation of IREEs in the case of Deeply-Virtual kinematics can be found
in Ref.~\cite{eit}.} of  $T^{(\textbf{F})}_{q,g}$on $\mu$. As a result, the r.h.s. of Eq.~(\ref{eqtktf}) are zeros.
A general solution to
Eq.~(\ref{eqtktf}) is

\begin{equation}\label{tfgen}
T^{(\textbf{F})}_q = \Psi_q \left((\rho - y),(\rho - z)\right),~~T^{(\textbf{F})}_g = \Psi_g \left((\rho - y),(\rho - z)\right),
\end{equation}
with $\Psi_q$ and $\Psi_g$ being arbitrary functions. In order to specify them
we use the following matching:
Amplitudes $T^{(\textbf{F})}_{q,g}$ coincide with amplitudes  $T^{(\textbf{E})}_{q,g}$
 on the border surface between regions \textbf{E} and \textbf{F},
where $\rho = y + z$:

\begin{equation}\label{matchef}
T^{(\textbf{F})}_q (y,z) = \bar{T}^{(\textbf{E})}_q (y,z), ~~T^{(\textbf{F})}_g (y,z) = \bar{T}^{(\textbf{E})}_g (y,z),
\end{equation}
with the new amplitudes $\bar{T}^{(\textbf{E})}_{q,g}$ defined as follows:

\begin{equation}\label{bart}
\bar{T}^{(\textbf{E})}_q (y,z) = T^{(\textbf{E})}_q (\rho,y,z)|_{\rho = y + z},~~
\bar{T}^{(\textbf{E})}_g (y,z) = T^{(\textbf{E})}_g (\rho,y,z)|_{\rho = y + z}.
\end{equation}

According to Eq.~(\ref{tfgen}) the expression for $T^{(\textbf{F})}_q, T^{(\textbf{F})}_g$ in the whole region \textbf{F}
can be obtained from amplitudes $\bar{T}^{(\textbf{E})}_{q,g}$ by the simple change of their arguments:

\begin{equation}\label{tf}
T^{(\textbf{F})}_q (\rho, y, z) = \bar{T}^{(\textbf{E})}_q (\rho -y, \rho - z),~~~
T^{(\textbf{F})}_g (\rho, y, z) = \bar{T}^{(\textbf{E})}_g (\rho -y, \rho - z).
\end{equation}

 Both $\rho - y$ and $\rho - z$ do not depend on the IR cut-off $\mu$, so amplitudes $T^{(\textbf{F})}_{q,g}$ are
 IR stable.

\subsection{Extension of amplitudes $T^{(\textbf{E,F})}_{q,g}$ to the region of small $Q^2$}

Extension of $T^{(\textbf{E})}_{q,g}$ to the small-$Q^2$ region can be done with the shift of
$Q^2$ like it was done in Collinear Factorization.
As a result, amplitudes $T^{(\textbf{E})}_{q,g}$ can be extended to the small-$Q^2$ region with replacement of $x, y$
in Eq.~(\ref{tkte}) by $\widetilde{x},\widetilde{y}$ defined in Eq.~(\ref{xytilde}).
A similar extension of $T^{(\textbf{F})}_{q,g}$ is impossible
because it would involve partons with virtualities $k^2_{\perp} > w$ which contradicts to Eq.~(\ref{wqk}).

\subsection{Extension of amplitudes $T^{(\textbf{E,F})}_{q,g}$ to Basic Factorization}

It follows from Ref.~\cite{bfact}  that extension of
 Compton amplitudes $T^{(\textbf{E})}_{q,g}$ and $T^{(\textbf{F})}_{q,p}$
defined in Eqs.~(\ref{tkte},\ref{tf}) to Basic factorization can be done with the very simple replacement:
it is necessary and sufficient to replace $k^2_{\perp}$ by $|k^2|$ in Eqs.~(\ref{tkte},\ref{tf}).

\subsection{Off-shell parton-parton amplitudes $H_{rr'}$}

Expressions
for Compton amplitudes $T^{(KT)}_{q,g}$ in Eqs.~(\ref{tkte},\ref{tf}) include off-shell Mellin amplitudes $H_{rr'}$. Below we calculate them in
DLA. They stem from the Mellin transform for off-shell parton-parton amplitudes $\widetilde{A}_{rr'}$:

\begin{equation}\label{mellinatilde}
\widetilde{A}_{rr'} (w,z) = \int_{- \imath \infty}^{\imath \infty} \frac{d \omega}{2 \pi \imath} e^{\omega \rho}
\xi^{(\omega)} \widetilde{f}_{rr'} (\omega,z)
\end{equation}
and the following definition (cf. (\ref{fh})):

\begin{equation}\label{ftildeh}
H_{rr'} (\omega,z) = \frac{1}{8 \pi^2} \widetilde{f}_{rr'} (\omega,z).
\end{equation}

Amplitudes $H_{rr'} (\omega,z)$ can also be found with constructing and solving appropriate IREEs.
The IREEs for the off-shell $H_{rr'}$  are quite similar to  Eq.~(\ref{eqphikt}):

\begin{eqnarray}\label{eqhkt}
\frac{\partial}{\partial z} H_{qq} + \omega H_{qq} &=& h_{qq} H_{qq} + h_{qg} H_{gq},
\\ \nonumber
\frac{\partial}{\partial z} H_{gq} + \omega H_{gq} &=& h_{gq} H_{qq} + h_{gg} H_{gq},
\\ \nonumber
\frac{\partial}{\partial z} H_{qg}  + \omega H_{qg} &=&  h_{qq} H_{qg} + h_{qg} H_{gg},
\\ \nonumber
\frac{\partial}{\partial z} H_{gg} + \omega H_{gg} &=& h_{gq} H_{qg} +  h_{gg} H_{gg},
\end{eqnarray}
where l.h.s. of each equation corresponds to applying operator $-\mu^2 d/d \mu^2$ to Eq.~(\ref{mellinatilde})
while each r.h.s involves convolutions of $H_{rr'}$ and $h_{rr'}$.
Specifying the general solution to Eq.~(\ref{eqhkt}) should be done with using the matching:

\begin{equation}\label{matchhkt}
H_{rr'}|_{z = 0} = h_{rr'}.
\end{equation}

Solving (\ref{eqhkt}) and using the matching (\ref{matchhkt}), we obtain the following expressions for $H_{rr'}$:

\begin{eqnarray}\label{hkt}
H_{qq}  &=& e^{- \omega z} \left[C_1 e^{\Omega_{(+)} z} + C_2 e^{\Omega_{(-)} z}\right],
\\ \nonumber
H_{gq}  &=& e^{- \omega z} \left[\frac{h_{qq} - \Omega_{(+)}}{h_{qg}}~ C_1 e^{\Omega_{(+)} z} +
\frac{h_{qq} - \Omega_{(-)}}{h_{qg}}~C_2 e^{\Omega_{(-)} z}\right],
\\ \nonumber
H_{gg}  &=& e^{- \omega z} \left[C'_1 e^{\Omega_{(+)} z} + C'_2 e^{\Omega_{(-)} z}\right],
\\ \nonumber
H_{qg}  &=& e^{- \omega z} \left[\frac{h_{gg} - \Omega_{(+)}}{h_{gq}}~ C'_1 e^{\Omega_{(+)} z} +
\frac{h_{gg} - \Omega_{(-)}}{h_{gq}}~C'_2 e^{\Omega_{(-)} z}\right].
\end{eqnarray}

Explicit expressions for the terms $\Omega_{(\pm)}$ are presented in Eq.~(\ref{omegapm}) while $C_{1,2}$ and $C'_{1,2}$ are
defined in Eq.~(\ref{ccprime}). The overall factor $e^{- \omega z}$ in Eq.~(\ref{hkt}) converts the IR-dependent Mellin factor
$(w/\mu^2)^{\omega}$ of Eq.~(\ref{mellinatilde}) in the IR-stable factor $\left(w/k^2_{\perp}\right)^{\omega}$.

\section{Photoproduction amplitudes}

It follows from Eqs.~(\ref{fact},\ref{afact}) that the perturbative components $A_{\gamma q}, A_{\gamma g}$ of the photoproduction
are related to the perturbative Compton amplitudes $T_{q,g}$ in a simple manner:

\begin{equation}\label{agammatqg}
A_{\gamma q} = T_q|_{q^2 = 0},~~~A_{\gamma g} = T_g|_{q^2 = 0}.
\end{equation}

Eq.~(\ref{agammatqg}) holds in any form of QCD factorization but expressions for the photoproduction amplitudes
are different in different forms of factorizations. We start with obtaining them in Collinear Factorization.

\subsection{Photoproduction amplitudes in Collinear kinematics}

According to  Eq.~(\ref{agammatqg}), putting $q^2 = 0$ in Eqs.~(\ref{tqgcolb}, \ref{fqgcolb}) should
yield $A^{(col)}_{\gamma q}$ and $A^{(col)}_{\gamma q}$ in DLA.
However, such procedure cannot be done in the straightforward way because the Mellin amplitudes $F^{(col)}_{q,g}$ in (\ref{fqgcolb})
were obtained in Ref.~\cite{etf1}  for
$Q^2 \gg \mu^2$ only and they cannot be used at $Q^2 < \mu^2$. In order to describe $A^{(col)}_{q,g}$ at low $Q^2$, including $Q^2 = 0$, we
use the prescription in the Sect.~II and replace $x,y$ in Eqs.~(\ref{tqgcolb}, \ref{fqgcolb}) by $\widetilde{x},\widetilde{y}$
defined in Eq.~(\ref{xytilde}). Then, putting $Q^2 = 0$ in $\widetilde{x},\widetilde{y}$, we obtain

\begin{equation}\label{xyzero}
\widetilde{x}|_{q^2 = 0} \equiv  x_0 = \mu^2/w,~~~\widetilde{y}|_{q^2 = 0} \equiv y_0 = 0.
\end{equation}

Replacement of $\widetilde{x},\widetilde{y}$ by $x_0, y_0$ in Eqs.~(\ref{tqgcolb}, \ref{fqgcolb})
allows us to obtain photoproduction amplitudes in DLA in Collinear Factorization:
Combining Eqs.~ (\ref{tqgcolb},\ref{fqgcolb}) and (\ref{xyzero}), we obtain:

\begin{eqnarray}\label{agammacolb}
A^{(col)}_{\gamma q} &=& a_{\gamma q} \int_{-\imath \infty}^{\imath \infty} \frac{d \omega}{2 \imath \pi} \left(x_0/ \beta\right)^{-\omega}
\xi^{(+)} (\omega)  f^{(col)}_{\gamma q} (\omega),
%\frac{ (\omega - h_{gg})}{G(\omega)},
\\ \nonumber
A^{(col)}_{\gamma g}  &=&  a_{\gamma q} \int_{-\imath \infty}^{\imath \infty} \frac{d \omega}{2 \imath \pi} \left(x_0/ \beta\right)^{\omega} \xi^{(+)} (\omega) %\frac{ h_{qg}}{G(\omega)}
f^{(col)}_{\gamma g} (\omega),
\end{eqnarray}
with

\begin{eqnarray}\label{fgammacolb}
f^{(col)}_{\gamma q} &=& a_{\gamma q} \frac{ (\omega - h_{gg})}{G(\omega)},
\\ \nonumber
f^{(col)}_{\gamma g}  &=&  a_{\gamma q} \frac{ h_{qg}}{G(\omega)},
\end{eqnarray}

where $a_{\gamma q}$ is  the averaged photon-quark coupling $a_{\gamma q} = \bar{e_q^2}$, $x_0 = \mu^2/w$
and
\begin{equation}\label{detg}
G = (\omega - h_{qq})(\omega - h_{gg})- h_{gg}h_{qg}.
\end{equation}

Accounting for non-DL radiative correction in $A^{(col)}_{\gamma q}, A^{(col)}_{\gamma g}$ can be done by the same way as it
was done in Sec.~IIB for the Compton amplitudes.

\subsection{Photoproduction amplitudes in $K_T$- Factorization }

DL contributions to the perturbative photoproduction amplitudes  $A^{(KT)}_{\gamma q}(s, k^2_{\perp})$
and $A^{(KT)}_{\gamma g}(s, k^2_{\perp})$
come from the region

\begin{equation}\label{wk}
 w \gg k^2_{\perp}.
 \end{equation}

We will use the Mellin transform for $A^{(KT)}_{\gamma q}$ and $A^{(KT)}_{\gamma g}$  in the following
form:

\begin{eqnarray}\label{mellinakt}
A^{(KT)}_{\gamma q} &=& \int_{-\imath \infty}^{\imath \infty} \frac{d \omega}{2 \imath \pi} \left( w/\mu^2\right)^{\omega}
\xi^{(+)} (\omega) F^{(KT)}_{\gamma q} (\omega, z),
\\ \nonumber
A^{(KT)}_{\gamma g} &=& \int_{-\imath \infty}^{\imath \infty} \frac{d \omega}{2 \imath \pi} \left(w/\mu^2\right)^{\omega}
\xi^{(+)} (\omega) F^{(KT)}_{\gamma g} (\omega, z).
\end{eqnarray}

IREE for the Mellin amplitudes  $F_{\gamma q} (\omega, z),F_{\gamma g} (\omega, z)$ are similar to Eq.~(\ref{eqhkt}):

\begin{eqnarray}\label{eqakt}
\frac{\partial}{\partial z} F^{(KT)}_{\gamma q} (\omega,z) + \omega F^{(KT)}_{\gamma q} (\omega,z)  = f^{(col)}_{\gamma q} (\omega) H_{qq} (\omega,z)
+ f^{(col)}_{\gamma g} (\omega) H_{gq} (\omega,z),
\\ \nonumber
\frac{\partial}{\partial z} F^{(KT)}_{\gamma g} (\omega,z) + \omega F^{(KT)}_{\gamma g} (\omega,z)  = f^{(col)}_{\gamma q} (\omega) H_{qg} (\omega,z)
+ f^{(col)}_{\gamma g} (\omega) H_{gg} (\omega,z).
\end{eqnarray}

Expressions for on-shell amplitudes $f^{(col)}_{\gamma q} $ and $f^{(col)}_{\gamma g} $ in the r.h.s. of Eq.~(\ref{eqakt}) are given by
Eq.~(\ref{fgammacolb}) while new off-shell parton-parton amplitudes $H_{rr'} (\omega,z)$ are unknown and should be specified. The general solution to
Eq.~(\ref{eqakt}) should be specified by the matching:

\begin{equation}\label{matcha}
A^{(KT)}_{\gamma q}(w,z) |_{z = 0} = A^{(col)}_{\gamma q} (w).
\end{equation}

Solving Eq.~(\ref{eqakt}) and using the matching of Eq.~(\ref{mellinakt}) yields expressions for
 $A^{(KT)}_{\gamma q}$ and $A^{(KT)}_{\gamma g}$:

\begin{eqnarray}\label{akt}
A^{(KT)}_{\gamma q} &=& \int_{-\imath \infty}^{\imath \infty} \frac{d \omega}{2 \imath \pi} \left( w/k^2_{\perp}\right)^{\omega}
\xi^{(+)} (\omega)
\left[f^{(col)}_{\gamma q}\left(1  +
%f^{(col)}_{\gamma q}
\int_0^z d u e^{ \omega u} H_{qq} (\omega,z')\right) +
f^{(col)}_{\gamma g} \int_0^z d u e^{ \omega u} H_{gq} (\omega,z')
\right],
\\ \nonumber
A^{(KT)}_{\gamma g} &=& \int_{-\imath \infty}^{\imath \infty} \frac{d \omega}{2 \imath \pi} \left( w/k^2_{\perp}\right)^{\omega}
\xi^{(+)} (\omega)
\left[
%f^{(col)}_{\gamma g} +
f^{(col)}_{\gamma q} \int_0^z d u e^{ \omega u} H_{qg} (\omega,z') +
f^{(col)}_{\gamma g}\left(1 +  \int_0^z d u e^{ \omega u} H_{gg} (\omega,z')\right)
\right].
\end{eqnarray}

We remind that expressions for amplitudes $f^{(col)}_{\gamma q}$ and $f^{(col)}_{\gamma g}$ can
are obtained in Eq.~(\ref{fgammacolb}) while $H_{rr'}$ are defined in Eq.~(\ref{hkt}).
To conclude this Sect. we notice that non-DL corrections can be incorporated in Eqs.~(\ref{agammacolb},
\ref{akt}) absolutely similarly to the prescription of Sect.~IIB. Transition from $K_T$ Factorization to
Basic one is done with replacement of $k^2_{\perp}$ by $|k^2|$.

\section{Conclusion}

In this paper we have studied first the perturbative amplitudes $T_{q,g}$ of elastic Compton scattering off quarks and gluons
and then perturbative components $A_{\gamma q},A_{\gamma g}$ of the photoproduction amplitudes.
$T_{q,g}$ were calculated in Ref.~\cite{etf1} in DLA at small $x$ and large photon virtualities $Q^2$ (i.e. in
kinematic region \textbf{B})
in the framework of Collinear Factorization . We converted these results in expressions
which can be used
at arbitrary $x$ and $Q^2$ in Collinear, $K_T$ and Basic Factorizations. Extension to Region \textbf{A} was done with
combining the total resummation of DL contributions and DGLAP description of $T_{q,g}$. By doing so we obtained
in Eq.~(\ref{tqgcolbtilde}) the interpolation expressions
reproducing $T_{q,g}$ in DLA at small $x$ and coinciding with the DGLAP description of $T_{q,g}$ when $x$ are not small.
Using the shift of Eq.~(\ref{shift}), we extended description $T_{q,g}$ to small $Q^2$. As a result, we arrived at
Eq.~(\ref{tqgcol}) to the expressions for $T_{q,g}$ in Collinear Factorization, which can be used at arbitrary $x$ and $Q^2$.

In contrast to Collinear Factorization,  $T_{q,g}$ are the essentially off-shell in $K_T$ Factorization and they  cannot be
described in DLA by a single expression valid at arbitrary values of virtualities $k^2_{\perp}$ of the external partons.
It made us  consider separately regions of moderate (Eq.~(\ref{mvkin})) and deep (Eq.~\ref{dvkin}) virtualities and
obtain expressions  Eqs.~(\ref{tkte}) and (\ref{tf}) for $T_{q,g}$ in those regions. We obtained them by constructing
appropriate IREEs and solving them.

After transition from Region \textbf{B} to the to low $Q^2$ Region \textbf{C} has been studied, we became able to obtain explicit
expressions (\ref{agammacolb}) and (\ref{akt}) for  the photoproduction amplitudes $A_{\gamma q}, A_{\gamma g}$  in Collinear and $K_T$
Factorizations respectively.

Small-$x$ asymptotics of amplitudes $T_{q,g}$ are of the Regge type with the same intercept in any form of QCD factorization.
Because of it we considered such asymptotics in Collinear Factorization (see Eq.~(\ref{asympt})) and discussed dependence
of its intercept on accuracy of the calculations: the higher is the accuracy, the lesser is the intercept. In other words,
the hard Pomeron becomes the soft one, when the accuracy grows. We think that further increasing the accuracy can lead to
vanishing supercritical Pomeron(s) and restoration of the Unitarity.

The next interesting point is that when $T_{q,g}$ are calculated in DLA, their
asymptotics depends on the single variable $\zeta = Q^2/x^2$ as shown in Eq.~(\ref{asymptzeta}). Neither DGLAP nor
BFKL cause such dependence.

It is important to use the Regge asymptotics within their applicability region, i.e. at
$x < x_{max}$, with  $x_{max}$ given by Eq.~(\ref{xmax}).
When $x > x_{max}$ the asymptotic is considerably less than the parent amplitudes,
so such amplitudes should be used instead of the asymptotics. Ignoring this point
leads to various misconceptions: for instance, appearance of
artificial/model (hard) Pomerons and spin-dependent Pomerons.

To conclude, let us notice that DL Pomeron can play an important role for description of various hadronic reactions
where Pomerons are used
and the diffractive DIS in the first place, see e.g. \cite{newman}.

\section{Acknowledgement}

We are grateful to S.I.~Alekhin, D.Yu.~Ivanov, G.I.~Lykasov, F.~Olness  and O.V.~Teryaev for useful communications.

\appendix

\section{Expressions for the parton-parton amplitudes}

\begin{eqnarray}\label{h}
&& h_{qq} = \frac{1}{2} \Big[ \omega - Z - \frac{b_{gg} -
b_{qq}}{Z}\Big],\qquad h_{qg} = \frac{b_{qg}}{Z}~, \\ \nonumber &&
h_{gg} = \frac{1}{2} \Big[ \omega - Z + \frac{b_{gg} -
b_{qq}}{Z}\Big],\qquad h_{gq} =\frac{b_{gq}}{Z}~,
\end{eqnarray}
where
\begin{equation}
\label{z1}
 Z = \frac{1}{\sqrt{2}}\sqrt{ Y + W
}~,
\end{equation}
with
\begin{equation}\label{y1}
Y = \omega^2 - 2(b_{qq} + b_{gg})
\end{equation}
and
\begin{equation}\label{w}
  W = \sqrt{(\omega^2 - 2(b_{qq} + b_{gg}))^2 - 4 (b_{qq} - b_{gg})^2 -
16b_{gq} b_{qg} }.
\end{equation}

Eqs.~(\ref{h},\ref{y1},\ref{w}) express $h_{rr'}$ through terms $b_{rr'}$.
The terms $b_{rr'}$ include the Born factors $a_{rr'}$ and contributions of non-ladder graphs $V_{rr'}$:
\begin{equation}\label{bik}
b_{rr'} = a_{rr'} + V_{rr'}.
\end{equation}

The Born factors are (see Ref.~\cite{egtg1sum} for detail):

\begin{equation}\label{app}
a_{qq} = \frac{A(\omega)C_F}{2\pi},~a_{qg} = \frac{A'(\omega)C_F}{\pi},~a_{gq} = -\frac{A'(\omega)n_f}{2 \pi}.
~a_{gg} = \frac{2N A(\omega)}{\pi},
\end{equation}
where $A$ and $A'$ stand for the running QCD couplings:

\begin{eqnarray}\label{a}
A = \frac{1}{b} \left[\frac{\eta}{\eta^2 + \pi^2} - \int_0^{\infty} \frac{d z e^{- \omega z}}{(z + \eta)^2 + \pi^2}\right],
%\\ \nonumber
A' = \frac{1}{b} \left[\frac{1}{\eta} - \int_0^{\infty} \frac{d z e^{- \omega z}}{(z + \eta)^2}\right],
\end{eqnarray}
with $\eta = \ln \left(\mu^2/\Lambda^2_{QCD}\right)$ and $b$ being the first coefficient of the Gell-Mann- Low function. When the running effects for the QCD coupling
are neglected,
$A(\omega)$ and $A'(\omega)$ are replaced by $\alpha_s$. The terms $V_{rr'}$ are represented in a similar albeit more involved way (see Ref.~\cite{egtg1sum} for detail):

\begin{equation}
\label{vik} V_{rr'} = \frac{m_{rr'}}{\pi^2} D(\omega)~,
\end{equation}
with
\begin{equation}
\label{mik} m_{qq} = \frac{C_F}{2 N}~,\quad m_{gg} = - 2N^2~,\quad
m_{gq} = n_f \frac{N}{2}~,\quad m_{qg} = - N C_F~,
\end{equation}
and
\begin{equation}
\label{d} D(\omega) = \frac{1}{2 b^2} \int_{0}^{\infty} d z
e^{- \omega z} \ln \big( (z + \eta)/\eta \big) \Big[
\frac{z + \eta}{(z + \eta)^2 + \pi^2} - \frac{1}{z +
\eta}\Big]~.
\end{equation}

\section{Explicit expressions for ingredients of Eq.~(\ref{fqgcolb})}

The factors $C^{(\pm)}_{q,g}$ and $\Omega_{(\pm)}$ of Eq.~(\ref{tqgcolb} were obtained in Ref.~\cite{etf1}.
We list them below. All of them are expressed though the parton-parton amplitudes $h_{rr'}$ of Appendix A.
The factors $\Omega_{(\pm)}$  are:

\begin{equation}\label{omegapm}
\Omega_{(\pm)} = \frac{1}{2} \left[ h_{gg} + h_{qq} \pm \sqrt{R}\right]
\end{equation}
and

\begin{equation}\label{r}
R = (h_{gg} + h_{qq})^2 - 4(h_{qq}h_{gg} - h_{qg}h_{gq}) = (h_{gg} - h_{qq})^2  + 4 h_{qg}h_{gq} .
\end{equation}

The factors $C_{(\pm)}$ are also expressed through the parton-parton amplitudes:

\begin{eqnarray}\label{cpmq}
C_q^{(+)} &=& a_{\gamma q}\frac{h_{qg}h_{gq} - (\omega - h_{gg})\left(h_{gg} - h_{qq} - \sqrt{R}\right)}{2 G \sqrt{R}},
\\ \nonumber
C_q^{(-)} &=& a_{\gamma q}\frac{ -h_{qg}h_{gq} + (\omega - h_{gg})\left(h_{gg} - h_{qq} + \sqrt{R}\right)}{2 G \sqrt{R}}.
\end{eqnarray}

\begin{eqnarray}\label{cpmg}
C_g^{(+)} &=& C_q^{(+)} \frac{h_{gg} - h_{qq} + \sqrt{R}}{2h_{qg}},
\\ \nonumber
C_g^{(-)} &=& C_q^{(-)} \frac{h_{gg} - h_{qq} - \sqrt{R}}{2h_{qg}}.
\end{eqnarray}

\section{Expressions for the factors $C_{1,2}$ and $C'_{1,2}$ of Eq.~(\ref{hkt}) }

The terms

\begin{eqnarray}\label{ccprime}
C_1 = \frac{2 h_{qg}h_{gq} - h^2_{qq} + h_{qq}h_{gg} + h_{qq}\sqrt{R}}{2 \sqrt{R}},
\\ \nonumber
C_2 = \frac{h^2_{qq} - h_{qq}h_{gg} - 2 h_{qg}h_{gq} + h_{qq}\sqrt{R}}{2 \sqrt{R}},
\\ \nonumber
C'_1 = \frac{2 h_{qg}h_{gq} - h^2_{gg} + h_{qq}h_{gg} + h_{gg}\sqrt{R}}{2 \sqrt{R}},
\\ \nonumber
C'_2 = \frac{h^2_{gg} - h_{qq}h_{gg} - 2 h_{qg}h_{gq} + h_{gg}\sqrt{R}}{2 \sqrt{R}},
\end{eqnarray}
where $R$ is given by Eq.~(\ref{r}).

\end{document}